\documentclass[preprint,amsmath,amssymb,aps,showkeys,showpacs]{revtex4}
\usepackage[colorlinks=true, pdfstartview=FitV, linkcolor=blue, citecolor=red, urlcolor=magenta]{hyperref}
\usepackage{graphicx}
\usepackage{latexsym}
\usepackage{amsmath}
\usepackage{amsfonts}
\usepackage{amssymb}
\usepackage{verbatim}
\usepackage{dcolumn}
\usepackage{amssymb}
\usepackage{bm}
\usepackage{float}
\usepackage{subfigure}
\usepackage[english]{babel}
\newcommand{\be}{\begin{equation}}
\newcommand{\ee}{\end{equation}}
\newcommand{\bea}{\begin{eqnarray}}
\newcommand{\eea}{\end{eqnarray}}

\begin{document}

\newcommand{\NITK}{
\affiliation{Department of Physics, National Institute of Technology Karnataka, Surathkal  575 025, India}
}

\newcommand{\IIT}{\affiliation{
Department of Physics, Indian Institute of Technology, Ropar, Rupnagar, Punjab 140 001, India
}}

\title{Microstructure of five-dimensional neutral Gauss-Bonnet black hole in anti-de Sitter spacetime via $P-V$ criticality}

\author{Naveena Kumara A.}
\email{naviphysics@gmail.com}
\NITK
\author{Ahmed Rizwan C.L.}
\email{ahmedrizwancl@gmail.com}
\NITK
\author{Kartheek Hegde}
\email{hegde.kartheek@gmail.com}
\NITK
\author{Md Sabir Ali}
\email{alimd.sabir3@gmail.com}
\IIT
\author{Ajith K.M.}
\email{ajith@nitk.ac.in}
\NITK

\begin{abstract}
In this article, we analytically investigate the microstructure of a five-dimensional neutral Gauss-Bonnet black hole, in the background of anti-de Sitter spacetime, using scalar curvature of the Ruppeiner geometry constructed via adiabatic compressibility. The microstructure details associated with the small-large black hole phase transition are probed in the parameter space of pressure and volume. The curvature scalar shows similar properties for both phases of the black hole, it diverges in the vicinity of critical point and approaches zero for extremal black holes.  We show that the dominant interaction among black hole molecules is attractive. This study also affirms that the nature of the microstructure interaction remains unchanged during the small-large black hole phase transition, even though the microstructures are different for both phases. 
\end{abstract}

\keywords{Black hole thermodynamics, five-dimensional neutral Gauss-Bonnet black hole, Extended phase space, van der Waals fluid, Ruppeiner geometry, Black hole microstructure}
%\pacs{}

\maketitle

%%%%%%%%%%%%%%%%%%%%%%%%%%%%%%%%%%%%%%%%%%%%%%%%%%%%%%%%%%%%%%%%%%%%%%%%%%%%%%%%%%%%%%%%%%%%%%%%%%%%%%%%%%%%%%%%%%%%%%%%%%%%%%%%%%%%%%%%%%%%%%%%%%%%%%%%%%%

\section{Introduction}
Recent developments in black hole chemistry suggest that the phase transition study can be used to probe the intriguing properties of black holes. The black hole phase transition study in AdS space is eased by the identification of the cosmological constant with the thermodynamic variable pressure \cite{Kastor:2009wy, Dolan:2011xt}. What followed from the studies is that the phase transition properties of certain AdS black holes are akin to conventional van der Waals system \cite{Kubiznak2012, Gunasekaran2012, Kubiznak:2016qmn}. Interestingly, numerous recent studies are focused on understanding the phenomenological properties of the black hole microstructure in AdS spacetime via the phase transition study \cite{Wei2015, Wei2019a, Wei2019b, Guo2019, Miao2017, Zangeneh2017, Wei:2019ctz, Kumara:2019xgt, Kumara:2020mvo, Xu:2019nnp, Chabab2018, Deng2017, Miao2019a, Chen2019, Du2019, Dehyadegari2017, Ghosh:2019pwy, Ghosh:2020kba, Yerra:2020oph, Wu:2020fij}. These efforts are inspired by the applications of the Ruppeiner geometry methods to the conventional thermodynamic systems \cite{Ruppeiner95, Janyszek_1990, Oshima_1999x, Mirza2008, PhysRevE.88.032123}. However, there is a significant difference in the approach in the context of black hole thermodynamics.  In a conventional statistical investigation of a thermal system, the macroscopic details can be constructed from the microscopic details. In a black hole system, the perspective is different, where the microscopic details are derived from the macroscopic thermodynamic knowledge \cite{Ruppeinerb2008}. The elegance of Ruppeiner geometry method is that the sign of the curvature scalar, constructed in appropriate parameter space, gives hints of the nature of microscopic interactions. A negative sign stands for attractive interaction, whereas the positive sign for the repulsive interaction. Moreover, the behaviour near the critical point signifies the validity of the parameter space chosen for the study. The divergence of the curvature scalar near the critical point of phase transition tells that the quantity is in turn related to the correlation length of the interacting constituents. This provides an insight on the connection between the curvature scalar and the microstructure of constituents.

Recently, a more fundamental approach in this regard in Ref. \cite{Wei2019a} ignited more interest among the researchers. The Ruppeiner geometry method was framed from the Boltzmann entropy formula by choosing the fluctuation coordinates as temperature and volume of the thermodynamic system. A normalised universal metric, thus constructed, was used to study the microstructure interaction.
When applied to a conventional van der Waals system this method gives expected results, where the dominant interaction among the fluid molecules is always attractive in all phases. Surprisingly charged black holes showed deviation from this, wherein a repulsive dominant interaction was also observed for certain physical conditions \cite{Wei2019a, Wei2019b, Kumara:2020ucr}. However this result is not universal for all black holes, the five-dimensional neutral Gauss-Bonnet AdS black holes resembles exactly to a van der Waals fluid \cite{Wei:2019ctz}. These contrasting results show that the black hole microstructure shows more rich properties than a conventional thermodynamic system.

The primary motivation for our research is due to the newly proposed Ruppeiner geometry method for studying black hole microstructure of charged AdS black hole \citep{Dehyadegari:2020ebz}. In which authors have proposed a new normalised curvature scalar, constructed via adiabatic compressibility. The parameter space coordinates are the pressure and the thermodynamic volume of the black hole. For charged black holes it is observed that strong repulsive interaction dominate among the constituents of near extremal small black holes, and the new curvature scalar diverges for the extremal black holes. We seek the microscopic properties of the five-dimensional Gauss-Bonnet black holes using this new method. The similarity or the difference of the result obtained will help us to understand the microstructures of the black hole in this modified gravity.

The paper is organised as follows. In the next section, we discuss the thermodynamics and the phase transition of the black hole. Then the microstructure study is carried out by constructing the Ruppeiner geometry using the fluctuation coordinates as pressure and volume  (section \ref{sectwo}). In the last section ( \ref{secthree}) we present our findings.

%%%%%%%%%%%%%%%%%%%%%%%%%%%%%%%%%%%%%%%%%%%%%%%%%%%%%%%%%%%%%%%%%%%%%%%%%%%%%%%%%%%%%%%%%%%%%%%%%%%%%%%%%%%%%%%%%%%%%%%%%%%%%%%%%%%%%%%%%%%%%%%%%%%%%%%%%%%

\section{Thermodynamics and Phase Transition of the Black Hole}
\label{secone}
\begin{figure}[t]
\centering
\includegraphics[scale=0.8]{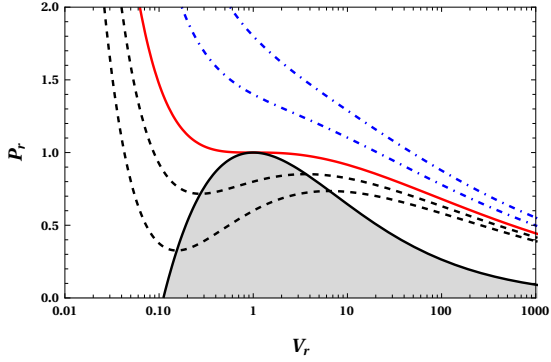}
\caption{$P-V$ isotherms of the five dimensional Gauss Bonnet AdS black hole.  The isotherms are shown in reduced parameters.  The shaded (grey) region below the solid black line corresponds to the unstable states.  (The $x$ axis is in the log scale).   }
\label{fig1}
\end{figure}

In this section we outline the extended phase space thermodynamics of A five dimensional neutral Gauss-Bonnet AdS black hole. We begin by considering the most general form of the action which describes a d-dimensional charged GB-AdS black hole is,
\begin{equation}
S=\int d^d x\sqrt{-g}\left( \frac{1}{16\pi G_d}\left(\mathcal{R}-2\Lambda+\alpha_{GB}\mathcal{L}_{GB}\right)-\mathcal{L}_{matter}\right)
\end{equation}
where the Lagrangian densities are given by,
\begin{align}
\mathcal{L}_{GB}&=\mathcal{R_{\mu\nu\gamma\delta}}\mathcal{R^{\mu\nu\gamma\delta}}-4\mathcal{R_{\mu\nu}}\mathcal{R^{\mu\nu}}+\mathcal{R}^2\\
\mathcal{L}_{matter}&=4\pi \mathcal{F_{\mu\nu}}\mathcal{F^{\mu\nu}}.
\end{align}
In the above exprssion $\alpha _{GB}$ is the Gauss-Bonnet coupling constant and  $\mathcal{F_{\mu\nu}}=\partial_\mu A_\nu-\partial_\nu A_\mu $ is the Maxwell field strength with the vector potential $A_{\mu}$. The spherically symmetric solution for the above action is given by \citep{Boulware:1985wk, Cai:2001dz, Wiltshire:1985us, Cvetic:2001bk} ,
\begin{equation}
ds^2=-f(r) dt^2+ f^{-1}(r) dr^2 + r^2\left(d\theta^2 +\sin^2\theta d\phi^2 +\cos^2 \theta d\Omega^2_{d-4}\right)
\end{equation}
with the corresponding metric function,
\begin{align}
f(r)=1+\frac{r^2}{2\alpha} \left(1-\sqrt{1+\frac{2\alpha}{d-2}\left( \frac{32\pi M}{\Sigma_{d-2}r^{d-1}}-\frac{4 Q^2}{\left(d-3\right)r^{2d-4}}+\frac{4\Lambda}{d-1}\right)}\right).
\end{align}
where the new parameter $\alpha$ is related to GB coupling constant as $\alpha=\left(d-3\right)\left(d-4\right)\alpha_{GB}$ and $\Sigma_{d-2}$ is the area of a unit sphere in $(d-2)$ dimensions. The parameters $M$ and $Q$ in the solutions correspond to the black hole mass and charge, respectively. We intend to study the extended thermodynamics of the black hole, where the cosmological constant is taken as pressure by the relation,
\begin{equation}
P=-\frac{\Lambda}{8\pi}.
\end{equation}
The $d$ dimensional charged solution takes the simple form in form in five dimension for neutral case. For $d=5$ and $Q=0$ we have,
\begin{equation}
f(r)=1+\frac{r^2}{2\alpha}\left(1-\sqrt{1+\frac{32 M\alpha}{3\pi r^4}-\frac{16P \pi \alpha}{3}}\right)
\end{equation}
The choice of this simple form is motivated by the fact that it exhibits a phase transition analogous to the van der Waals fluid, namely the small-large black hole transition. Also there exists analytical expression for the coexistence curve which characterises this phase transition. This enables one to study the phase transition and hence the associated properties of the black hole analytically. Doing so we try to understand the influence of the GB coupling parameter on the black hole phase structure and underlying microstruture. 
The position of the black hole event horizon $(r+)$ is governed by the largest positive real root of $f(r+) = 0$. Using that condition we obtain the  mass of the black hole, which reads
\begin{equation}
M= \frac{\pi}{8}\left( 4 P \pi r_h ^4 + 3 r_h^2 +3 \alpha \right).
\end{equation}
In the extended phase space this is treated as the enthalpy of the black hole system, $M=H$ \citep{Cai:2013qga}. The Hawking temperature at the event horizon is evaluated as,
\begin{equation}
T=\frac{f'(r_+)}{4\pi}=\frac{8\pi P r_h^4 + 3 r_h}{6 \pi r_h^2 +12 \pi \alpha}.
\end{equation}
Now we can write the first law and Smarr relation for the black hole,
\begin{align}
d H &= TdS + VdP + \mathcal{A}d \alpha\\
2 H &= 3 T S- 2 P V + 2 \mathcal{A} \alpha,
\end{align}
where $S$ and $V$ are the entropy and the thermodynamic volume of the black hole. The vriable $\mathcal{A}$ is conjugate to the GB coupling constant $\alpha$. These thermodynamic variables have the following form, which are consistent with the first law and Smarr reltion. 
\begin{eqnarray}
S= \frac{\pi^2 r_h}{2}\left(r_h^2 +6 \alpha \right)~~
\mathcal{A}=-\frac{\pi}{8}\left(\frac{32 \pi P r_h^4+ 9 r_h^2-6\alpha}{r_h^2+2\alpha} \right),~~V=\frac{\pi^2 r_h^4}{2}.
\end{eqnarray}
In addition to this we define the specific volume as $\mathit{v}=\frac{4}{3}r_h$. Our investigation of black hole microstructure is based on the fact the black hole entropy $S$ is function of volume $V$ for a fixed value of GB coupling constant $\alpha$, which is evident from the above expressions. This result we will use in the next section. To review the thermodynamics we write the state equation,
\begin{equation}
P = \frac{T}{\mathit{v}}-\frac{2}{3\pi \mathit{v}^2} + \frac{32 T \alpha}{9 \mathit{v}^3}
\end{equation}
The equation of state has $\alpha$ dependence, which shows the van der Waals like phase transition is depends on it. The phase transition of the black hole is between a small black hole phase and a large black hole phase. The behavior of isotherms in the P-V plane which features this transition is shown in fig. \ref{fig1}. 

The critical point of transition is calculated by using the constraints,
$\left( \partial_{\mathit{v}} P\right)_T= \left( \partial_{\mathit{v}, \mathit{v}} P\right)_T=0$. Which are given by \citep{Mo:2016sel, Wei:2019ctz},

\begin{equation}
P_c=\frac{1}{48 \pi \alpha},~~ T_c= \frac{1}{2\pi \sqrt{6 \alpha}},~~ V_c= 18 \pi^2 \alpha^2,~~ \mathit{v}_c= \sqrt{\frac{32 \alpha}{3}}.
\end{equation}
Using this we define the reduced thermodynamic quantities as,

\begin{equation}
P_r=\frac{P}{P_c},~~ T_r=\frac{T}{T_c},~~ \mathit{v}_r= \frac{\mathit{v}}{\mathit{v}_c},~~V_r=\frac{V}{V_c}.
\end{equation}
In the reduced parameter space the equation of state takes the following form,
\begin{equation}
P_r=\frac{3T_r}{\mathit{v}_r}-\frac{3}{\mathit{v_r}^2}+ \frac{T_r}{\mathit{v}_r^3}
\end{equation}
It is clear that in the reduced parameter space the state equation is independent of the coupling constant $\alpha$. This inturn related to the fact that the black hole under consideration is a single-characteristic-parameter system \citep{Wei:2015ana}. For the temperature below the critical value $T_r<1$, the oscillating part of the isotherms in fig. \ref{fig1} stand for the unstable system which has a negative isothermal compressibility,
\begin{equation}
    \kappa _T=-\frac{1}{V}\left(\frac{\partial V}{\partial P}\right)_T<0
\end{equation}
These unstable regions are removed using the Maxweel equal area construction, $\oint VdP =0$. The construction signifies a first order phase transition of the system. However the equal area law construction can be carried out in $P-V$ plane, not in the $P-\mathit{v}$ plane. For that we use the equation of state in terms of the thermodynamic volume,
\begin{equation}
P= \frac{3\pi^{3/2} \alpha T}{2 \left(2 V\right)^{3/4}}+\frac{3 \sqrt{\pi}T}{4 \left(2 V\right)^{1/4}}-\frac{3}{8\sqrt{2}V}
\end{equation}
or in the reduced parameter space,
\begin{equation}
P_r= \frac{T_r}{V_r^{3/4}}-\frac{3}{\sqrt{V_r}}+ \frac{3 T_r}{V_r^{1/4}}.
\end{equation}
Equivalently one can construct the Maxwell equal area law in $T-S$ plane \citep{Spallucci:2013osa}. The coexistence curve hence obtained determines the small-large black hole transition region. The coexistence curve can also be obtained by other means, i.e., from the Gibbs free energy method. For the five dimensional neutral Gauss Bonnet black hole we have the following forms of coexistence curve in different planes \citep{Mo:2016sel, Wei:2019ctz},

\begin{equation}
P_r= \frac{1}{2}\left(3-\sqrt{9-8T_r^2}\right)
\end{equation}

 \begin{equation}
 T_r=\frac{3V_r^{/4}(1+\sqrt{V_r})}{1+4\sqrt{
 V_r}+ V_r}
 \end{equation}

The existence of these analytic solutions enables us to examine the microstructure of the black hole analytically, which we will perform in the next section.

%%%%%%%%%%%%%%%%%%%%%%%%%%%%%%%%%%%%%%%%%%%%%%%%%%%%%%%%%%%%%%%%%%%%%%%%%%%%%%%%%%%%%%%%%%%%%%%%%%%%%%%%%%%%%%%%%%%%%%%%%%%%%%%%%%%%%%%%%%%%%%%%%%%%%%%%%%%

\section{Ruppeiner Geometry and Microstructure of Black Hole}
\label{sectwo}
In this section, we probe the microstructure of five-dimensional Gauss-Bonnet black hole using Ruppeiner geometry constructed in the parameter space of fluctuation coordinates. Basically, this naive idea is developed from Gaussian thermodynamic fluctuation theory \citep{Ruppeiner95}. In this approach, a scalar curvature is constructed from a line element, which is the measure of the distance between two neighbouring fluctuation states of the thermodynamic system. The sign of the curvature scalar tells the nature of the interaction between the constituents of the thermal system, a negative sign for attractive and a positive sign for repulsive interaction. Vanishing curvature scalar corresponds to no interaction in this picture. This correspondence is adopted from the applications of Ruppeiner geometry in conventional thermodynamic systems \citep{Janyszek_1990, Oshima_1999x}. 

\begin{figure}[t]
\centering
\subfigure[ref1][]{\includegraphics[scale=0.8]{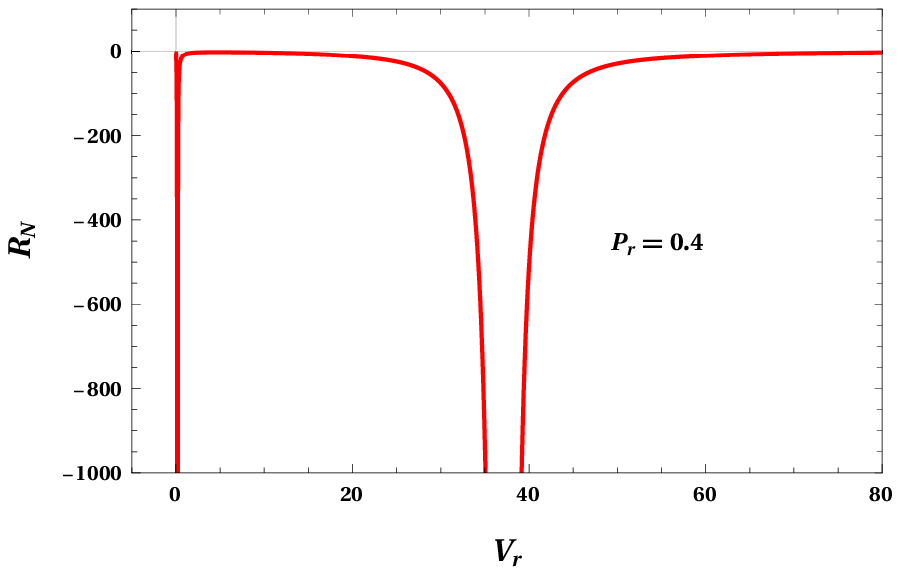}\label{RNVR1}}
\qquad
\subfigure[ref2][]{\includegraphics[scale=0.8]{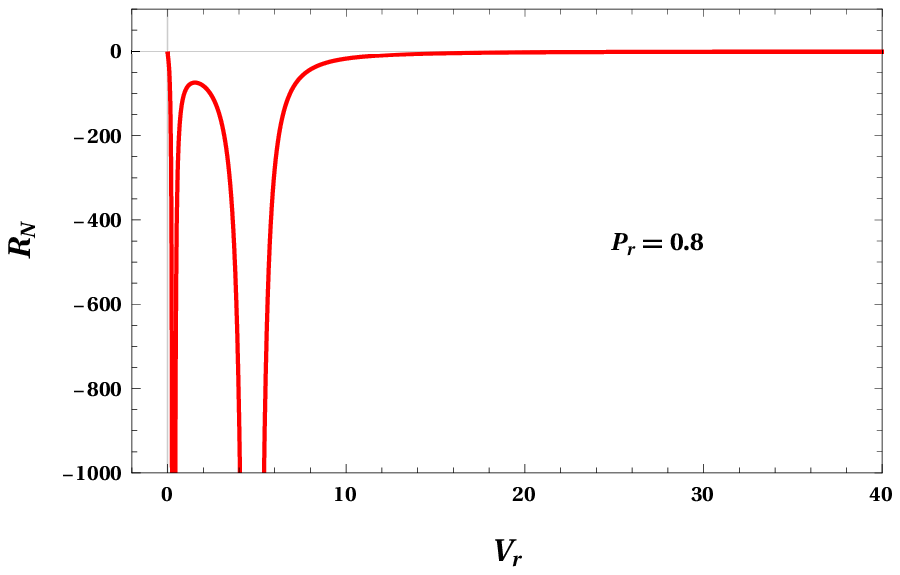}\label{RNVR2}}

\subfigure[ref1][]{\includegraphics[scale=0.8]{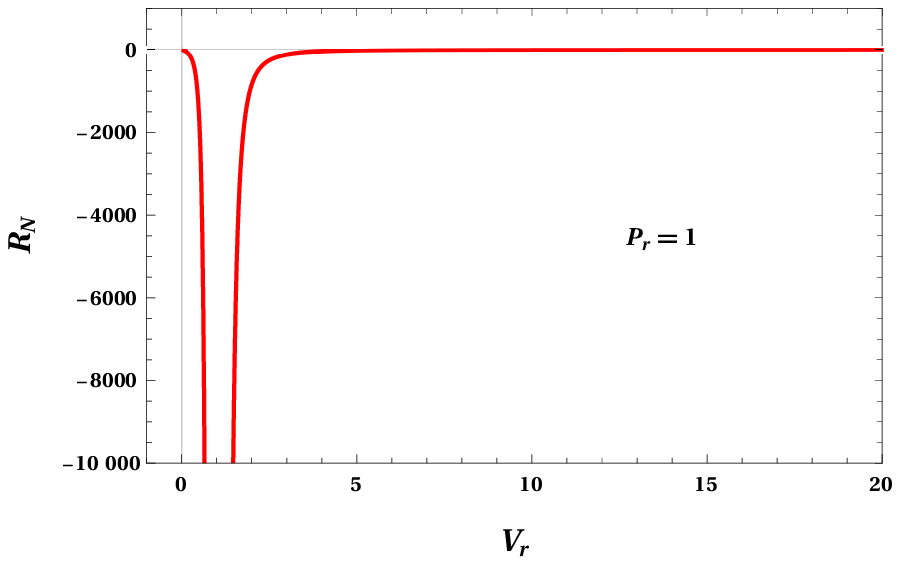}\label{RNVR3}}
\qquad
\subfigure[ref1][]{\includegraphics[scale=0.8]{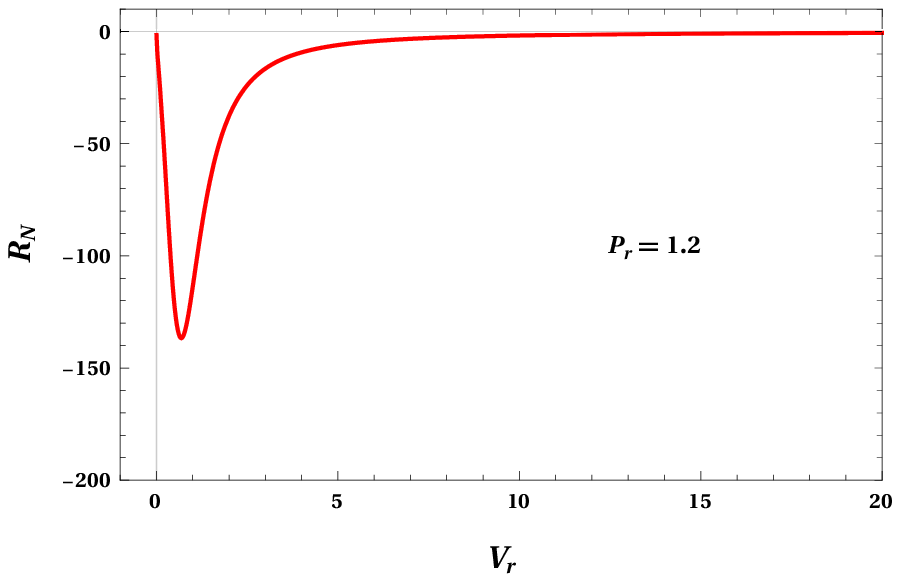}\label{RNVR4}}
\caption{The behaviour of the normalised curvature scalar $R_N$ against the reduced volume $V_r$ at constant pressure. }
\label{RN}
\end{figure}

In a recent work the microstructure of the charged AdS black hole is investigated by using a new fluctuation metric in the parameter space of pressure and temperature. It is shown that the fluctuation metric in the $(S,P)$ coordinates can be translated to $(P,V)$ coordinates, as the entropy of the black hole depends only on volume $V$. The line element in this parameter space of $(S,P)$ has the form \cite{Dehyadegari:2020ebz},
\begin{equation}
dl^2=\frac{1}{C_P}dS^2+\frac{V }{T}\kappa _S dP^2.
\label{line}
\end{equation}
Where the heat capacity at constant pressure $C_P$ and adiabatic compressibility $\kappa _S$ are given by,
\begin{equation}
  C_P=T\left( \frac{\partial S}{\partial T} \right) _P \qquad , \qquad \kappa_S=-\frac{1}{V}\left( \frac{\partial V}{\partial P} \right) _S.
\end{equation}
For neutral five dimensional Gauss Bonnet black hole we can write Eq. \ref{line} as,
\begin{equation}
dl^2=\frac{1}{C_P} \left( \frac{9 \pi  \left(2\times 2^{1/4} \pi  \alpha +2^{3/4} \sqrt{V}\right)^2}{64 V^{3/2}} \right) dV^2+\frac{V }{T}\kappa _S dP^2.
\label{line2}
\end{equation}
Now we have pressure and volume as the fluctuation variable, so that we can construct the thermodynamic geometry in $P-V$ plane. The adiabatic compressibility $\kappa _S$ is a vanishing quantity for a black hole, like the heat capacity at constant volume, $C_V$. To avoid the pathologies associated with the curvature a scalar due to this, a normalised thermodynamic scalar is proposed by Amin Dehyadegari et. al., \citep{Dehyadegari:2020ebz} as,

\begin{equation}
R_N=\kappa_S R.
\end{equation}
By the direct application of the definition of curvature scalar in Riemannian geometry to Eq. (\ref{line2}) we obtain normalised Ruppeiner scalar $R_N$ for the five dimensional neutral Gauss Bonnet black hole. In terms of the reduced parameters $P_r$ and $V_r$ it reads,

\begin{equation}
R_N=\frac{32 \sqrt{\frac{2}{3}} {V_r}^{1/4} \left(3 P_r  V_r+3 (P_r -1) \sqrt{V_r}+5\right)}{\pi  \left(P_r  \sqrt{V_r}+3\right) \left(P_r V_r+(P_r -3) \sqrt{V_r}+1\right)^2}.
\label{RNeqn}
\end{equation}
As in the case of charged AdS black hole, where no charge dependence, the normalised curvature scalar $R_N$ is independent of the Gauss Bonnet coupling parameter $\alpha$. The Eq. (\ref{line}) also gives the same expression for $R_N$, i.e., via in the $(S,P)$ parameter space. The functional behaviour of $R_N$ with reduced volume $V_r$ for a fixed pressure is studied in fig (\ref{RN}). 

The behaviour is similar to that of the normalised Ruppeiner scalar in $(T,V)$ parameter space \citep{Wei:2019ctz, Kumara:2020ucr}.  For $P_r<1$, below critical pressure, $R_N$ has two negative divergences. Those two divergences come nearer as the pressure increases and merge at $V_r=1$ for $P_r=1$. For the pressure above the critical value, $P>P_r$, there is no divergence for $R_N$. Another important feature to observe from these figures is that the curvature scalar $R_N$ is always positive, which is the indication that the dominant interaction in the black hole microstructure is always attractive. 

In fact we can study the overall behaviour of curvature scalar in $P_r-V_r$ plane. The divergences of $R_N$ in the $P_r-V_r$ plane can be obtained from Eq. (\ref{RNeqn}) as,
\begin{equation}
    P_{div}=\frac{3 \sqrt{\text{Vr}}-1}{\text{Vr}+\sqrt{\text{Vr}}},
    \label{diveqn}
\end{equation}
where we have taken only the positive root. We also obtain the curve along which the curvature scalar $R_N$ vanishes,
\begin{equation}
    P_0=\frac{3 \sqrt{\text{Vr}}-5}{3 \left(\sqrt{\text{Vr}}+1\right) \sqrt{\text{Vr}}}.
    \label{vanisheqn}
\end{equation}

\begin{figure}[H]
\centering
\subfigure[ref2][]{\includegraphics[scale=0.8]{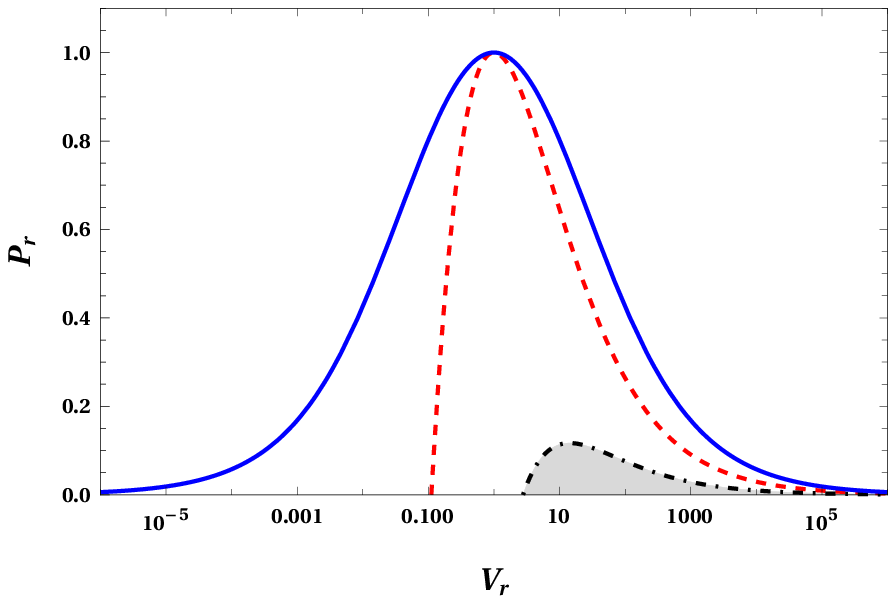}\label{Sign}}
\qquad
\subfigure[ref1][]{\includegraphics[scale=0.8]{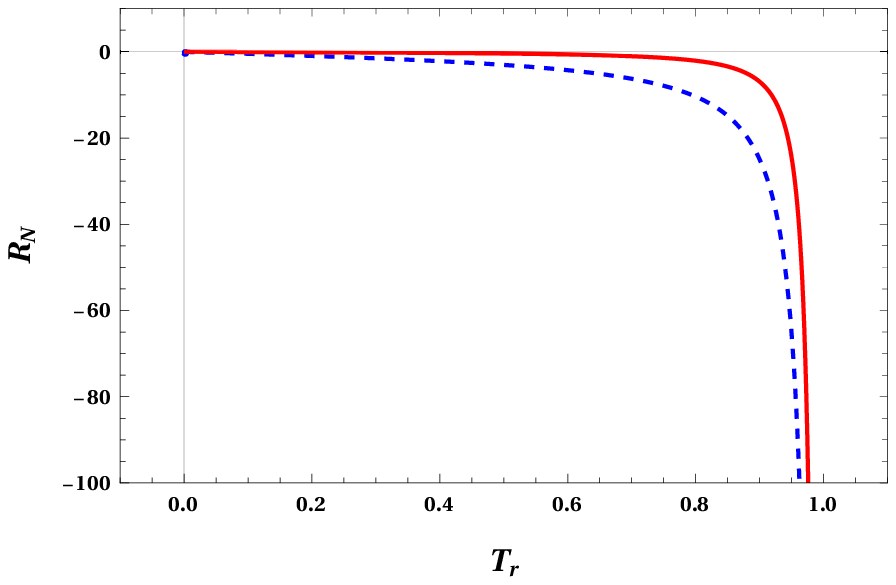}\label{RNTR}}
\caption{ \ref{Sign}:  The vanishing curve (red dashed line) and diverging curve (black dotdashed line) of $R_N$ along with the coexistence curve (blue solid line) . The shaded region (grey) corresponds to positive $R_N$, otherwise $R_N$ is negative. The $x$ axis is in log scale. \ref{RNTR}: The behaviour of normalised curvature scalar $R_N$ along the coexistence line. The red (solid) line and  blue (dashed) line corresponds to large black hole and small black hole, respectively.}
\end{figure}

This is the equation which governs the change in the interaction between the black hole molecules from attractive to repulsive and vice versa, provided the conditions are favourable for stability. To understand this, we depict Eq. (\ref{diveqn}) and Eq. (\ref{vanisheqn}) along with the coexistence curve in fig. \ref{Sign}. The shaded regions correspond to the positive sign of $R_N$ and hence the underlying repulsive interactions. In the remaining regions, $R_N$ takes negative values, implying the dominant attractive interactions. It is clear from fig. \ref{Sign} that the diverging curve and the vanishing curve are both under the coexistence curve. The equation of state does not hold under the coexistence curve, which is the region of coexistence of small and large black holes. There this region is excluded as the system is not stable here. Therefore, for five-dimensional neutral Gauss-Bonnet black hole, all the physically meaningful stable regions give a negative sign for $R_N$ always. This implies that there exist only dominant attractive interactions among the constituents of five-dimensional Gauss-Bonnet AdS black hole. This result is different from that of charged AdS black hole where repulsive interaction can exist for small black hole phase \cite{Dehyadegari:2020ebz, Wei2019a, Wei2019b}. These results are consistent with the earlier observation on the five-dimensional Gauss-Bonnet AdS black hole microstructure \citep{Wei:2019ctz}.

Finally, we consider the behaviour of the normalised curvature scalar $R_N$ along the coexistence curve (fig. \ref{RNTR}). Both the small black hole and the large black hole branches of $R_N$ diverges to $-\infty$ in the vicinity of the critical point and approaches zero for the extremal cases. The negative sign of $R_N$ for all parameter space affirms that there exists only dominant attractive interaction in the black hole microstructure. The interaction vanishes as the $T_r\rightarrow 0$ whereas the attractive interaction becomes stronger as $T_r\rightarrow T_c$. The diverging nature of $R_N$ near the critical point is similar to that of van der Waals case. From the analysis, we conclude that the dominant interaction of five-dimensional Gauss-Bonnet black hole is similar to van der interaction, where it is always attractive.

%%%%%%%%%%%%%%%%%%%%%%%%%%%%%%%%%%%%%%%%%%%%%%%%%%%%%%%%%%%%%%%%%%%%%%%%%%%%%%%%%%%%%%%%%%%%%%%%%%%%%%%%%%%%%%%%%%%%%%%%%%%%%%%%%%%%%%%%%%%%%%%%%%%%%%%%%%%

\section{Discussions}
\label{secthree}

In this article, we have probed the microstructure of five-dimensional neutral Gauss-Bonnet AdS black hole by studying its phase transition. We have adopted a new definition for the scalar curvature of the Ruppeiner geometry, constructed via adiabatic compressibility, which recently appeared in Ref. \citep{Dehyadegari:2020ebz}. The fluctuation coordinates in this description are the pressure and the volume of the black hole. The microscopic properties are understood by looking at the behaviour of this new normalised curvature scalar along the coexistence line and in the neighbourhood of the critical point. The study is carried out analytically. Two main findings from our investigation is that (i) the normalised curvature scalar $R_N=\kappa _S R$ is always negative for all physically meaningful regions in the parameter space, which implies an attractive interaction among the black hole molecules, (ii) in the vicinity of the critical point, $R_N$ diverges to negative infinity for both small black hole and large black hole phases. The similarity in the properties of $R_N$ for two phases of the black hole shows that, unlike charged AdS black holes, the nature of the interaction between black hole molecules is unchanged in Gauss-Bonnet AdS gravity during the phase transition. Also, the like results for microstructure in different parameter space, compared to the previous results in the literature \citep{Wei:2019ctz},  affirms that the nature of the interaction between microstructures is eternal to the system. We believe that there is more to explore in the black hole microstructure, and our study enhances the utility of the available tools to probe those properties.

%%%%%%%%%%%%%%%%%%%%%%%%%%%%%%%%%%%%%%%%%%%%%%%%%%%%%%%%%%%%%%%%%%%%%%%%%%%%%%%%%%%%%%%%%%%%%%%%%%%%%%%%%%%%%%%%%%%%%%%%%%%%%%%%%%%%%%%%%%%%%%%%%%%%%%%%%%%

\acknowledgments
Authors N.K.A., A.R.C.L. and K.H. would like to thank U.G.C. Govt. of India for financial assistance under UGC-NET-SRF scheme.

%%%%%%%%%%%%%%%%%%%%%%%%%%%%%%%%%%%%%%%%%%%%%%%%%%%%%%%%%%%%%%%%%%%%%%%%%%%%%%%%%%%%%%%%%%%%%%%%%%%%%%%%%%%%%%%%%%%%%%%%%%%%%%%%%%%%%%%%%%%%%%%%%%%%%%%%%%%

  \bibliography{BibTex}

\end{document}